\documentclass[12pt]{article}

\begin{document}

\sloppy
\begin{flushright}{SIT-HEP/TM-19}
\end{flushright}
\vskip 1.5 truecm
\centerline{\large{\bf Q ball inflation}}
\vskip .75 truecm
\centerline{\bf Tomohiro Matsuda
\footnote{matsuda@sit.ac.jp}}
\vskip .4 truecm
\centerline {\it Laboratory of Physics, Saitama Institute of
 Technology,}
\centerline {\it Fusaiji, Okabe-machi, Saitama 369-0293, 
Japan}
\vskip 1. truecm
\makeatletter
\@addtoreset{equation}{section}
\def\theequation{\thesection.\arabic{equation}}
\makeatother
\vskip 1. truecm

\begin{abstract}
\hspace*{\parindent}
We show that inflation can occur in the core of a
Q-ball.
\end{abstract}

\newpage
\section{Introduction}
\hspace*{\parindent}
In spite of the great success in the quantum field theory, there is still no
consistent scenario in which the quantum gravity is included.
The most promising scenario in this direction would be the string
theory, where the consistency is ensured by the requirement of
additional dimensions and supersymmetry.
Supersymmetry is sometimes used to explain the large hierarchy
between the scales of Grand Unified Theory(GUT) and the Standard
Model(SM).
Unlike the standard model, the Supersymmetric
Standard Model(SSM) contains many flat directions in its scalar sector.
The flat directions might appear from the moduli fields that
parametrizes the compactified
space of the string theory.
Moreover, considering the brane extensions of the string theory,
flat potential might appear from the potential for the 
distances between branes, or simply
for the positions of branes in the compactified
space.\footnote{Some cosmological implications and defect configurations
of the flat directions are
discussed in ref.\cite{matsuda_yukawa, incidental_matsuda}.}
In the string theory, initially the sizes of extra dimensions had been
assumed to be as small as the Planck mass, but later it has been
observed that there is no reason 
to believe such a tiny compactification radius\cite{Extra_1}.
The idea of the large extra dimension may solve the hierarchy problem.
Denoting the volume of the $n$-dimensional compact space by $V_n$,
the observed Planck mass is obtained by the relation $M_p^2=M^{n+2}_{*}V_n$,
where $M_{*}$ denotes the fundamental scale of gravity.
If one assumes more than two extra dimensions, $M_{*}$ may be
close to the TeV scale without conflicting any observable
bounds.
In this scenario the standard model fields are expected to be localized
on a wall-like structure and the graviton propagates in the bulk.
The most natural embedding of this picture in the string theory context 
is realized by a brane construction.\footnote{Constructing successful
models for inflation with a low fundamental scale is still an interesting
problem\cite{low_inflation, matsuda_nontach}.
Baryogenesis and inflation in models with a low fundamental scale are
discussed in \cite{low_baryo}.
Affleck-Dine baryogenesis\cite{AD} in such scenarios are discussed in
ref.\cite{low_AD, ADafterThermal}. 
We think constructing models of particle cosmology with large
extra dimensions is very important since we are expecting that future
cosmological observations would determine the fundamental
scale of the underlying theory.}

On the other hand, we know historically that the characteristic features of
the phenomonological models are revealed by discussing their cosmological
evolutions. 
For example, so called the Polonyi
problem for the flat directions of the supersymmetric models has been
discussed by many authors\cite{polonyi}. 
If the historical knowledge still applies to the above brane-extended flat
directions, 
it should be important to seek other options, since the alternative
approaches might reveal the 
underlying problems of the model or show us the novel approach to a new
scenario of cosmology.

From the above point of view, we will consider Q-balls and show that
inflation may start inside a Q-ball.
It is well known that a variety of models possess flat directions
that may lead to non-topological solitons of the Q-ball
type\cite{Q-ball, incidental_matsuda}.
On the other hand, it is sometimes discussed that topological defects 
other than the Q-balls, which might be produced in the early stage of
the Universe, would play 
important roles in particle cosmology.\footnote{Inflation from
topological defect is discussed in ref.\cite{topological_infla} for
conventional gravity and in ref.\cite{matsuda_defectinfla} for models
of the brane world. 
Constraints on hidden-sector walls from weak
inflation are discussed in ref.\cite{matsuda_weak_wall}.}
Here we briefly summarize the interesting differences between
conventional defect inflation and the Q-ball inflation.
\begin{itemize}
\item In conventional models of defect inflation, the typical parameters
      such as the mass scales or the width of the defect are the
      constants that do not change with time. In this case, the
      conditions for defect inflation to start are determined solely by
      the form of the potential that induces defect configuration. On
      the other hand, the typical parameters of the Q-balls will be
      determined by their charges that may evolve with time by absorbing
      other Q-balls.
\item Q-balls will be produced during the oscillation after inflation. If
      the Q-balls decay safely, the reheating may be dominated by the
      ``surface reheating'' that was discussed in
      ref.\cite{surface-reheat}. On the other hand, if a produced
      Q-ball triggers inflation in its core, the second stage of
      inflation will start in the local area. Of course the
      second inflation might produce another Q-balls again, which might
      induce problematic eternal inflation. 
\item The Q-ball inflation might start as the secondary weak inflation
      that induces the temporal swmall expansion of a local area. Thus, unlike
      conventional 
      defect inflation, the Q-ball inflation might be used to explain
      the bubble structure of the Universe. 
\end{itemize}

In section 2 we show a simple example of the Q-ball that appears on a
hybrid potential.
Although there had been no explicit argument on such an extension of the flat
potential of the Q-balls, it is easy to see that the extension of this
type is quite natural in many phenomonological models.
In section 3 we examine the conditions for the Q-ball inflation.
In section 4, we apply the results obtained in section 3 to a specific
model of surface reheating\cite{surface-reheat}. 
We stress here that the ideas we will show in this paper are quite simple
and generic.

\section{Q-balls from hybrid potential}
\hspace*{\parindent}
Here we consider a typical hybrid potential that has been used in models
of D-term inflation\cite{D-term}.
As usual, the trigger field $\phi$ is assumed to have a steep
potential, while the field $\sigma$ parameterizes a flat potential.
The explicit form of the effective potential is
\begin{equation}
\label{hybrid-potential}
V(\sigma)=M_0^4 \log \left(1+K_1 \frac{|\sigma|^2}{M_1^2}\right)+m_{3/2}^2
|\sigma|^2\left[1+ K_2 \log \left(\frac{|\sigma|^2}{M_2^2}\right)\right]
\end{equation}
where $m_{3/2}$ is the gravitino mass, which is obtained by using the
supersymmetry breaking scale $\Lambda_{SUSY}$, as
$m_{3/2}=\frac{\Lambda_{SUSY}^2}{M_p}$. 
The constant $K_1, K_2$ represent the renormalization factors at one-loop,
and $M_1, M_2$ are the renormalization scales.
We introduce a dimensionless constant $\eta \equiv M_0/\Lambda_{SUSY}$
for later convenience.
The second term dominates when 
$\sigma >\sigma_{c}$, where $\sigma_c$ is defined as 
\begin{eqnarray}
\sigma_c&\equiv&\frac{M_0^2}{m_{3/2}}\nonumber\\
&=& M_p\left( \frac{M_0}{\Lambda_{SUSY}} \right)^2\nonumber\\
&=& M_p\times \eta^2,
\end{eqnarray}
where the numerical factor is neglected.
For $\eta \gg 1$, $\sigma_c$ becomes much larger than the Planck scale 
($\sigma_c \gg M_p$), and the second term in
(\ref{hybrid-potential}) is always negligible.
On the other hand, for $\eta \ll 1$, one should consider two types of
Q-balls\cite{Kasuya-Kawasaki}.

Let us first consider the case where $\eta <1$.
As far as $\sigma < \sigma_{c}$, 
the potential is dominated by the first term in
eq.(\ref{hybrid-potential}) and the Q-balls will have the following
well-known properties\cite{Kasuya-Kawasaki,Dvali-flatQ}; 
\begin{eqnarray}
\label{Q-ball-prop-gauge}
r_Q \simeq \frac{Q^{1/4}}{M_0}, && \omega \simeq \frac{M_0}{Q^{1/4}}\nonumber\\
\sigma \simeq M_0 Q^{1/4}.&&
\end{eqnarray}
On the other hand, when $\sigma > \sigma_{c}$, the second term will
dominate and the Q-balls will have the following
properties\cite{Kasuya-Kawasaki},
\begin{eqnarray}
\label{Q-ball-prop-gra}
r_Q \simeq \frac{1}{\sqrt{|K_2|} m_{3/2}}, && \omega \simeq m_{3/2}\nonumber\\
\sigma \simeq |K|^{3/4} m_{3/2} Q^{1/2}.&&
\end{eqnarray}
As these results are well established, we use
(\ref{Q-ball-prop-gauge}) 
and (\ref{Q-ball-prop-gra}) hereafter to examine the conditions for the Q-ball
inflation.  
Because we are considering hybrid potential, we may assume 
$M_0 \gg \Lambda_{SUSY}$ as well as $M_0 \ll \Lambda_{SUSY}$.
If one assumes $M_0 \gg \Lambda_{SUSY}$, $\eta$ becomes much larger
than $O(1)$.
In this case, $\sigma$ is always much smaller than $\sigma_c$, which
means that one may safely assume that (\ref{Q-ball-prop-gauge}) is always 
satisfied for any $Q$.

In the next section we use the above results and examine the conditions
for the Q-ball inflation.
We then analyze the properties of the Q-ball inflation.

\section{Inflation from a Q-ball}
\hspace*{\parindent}
As in the conventional models of topological
inflation\cite{topological_infla}, the following conditions will be
required. 
\begin{itemize}
\item The radius of the Q-ball must become larger than
      the Hubble radius. Since the field $\sigma$ is trapped at a false
      vacuum within the Q-ball, inflation will start when the boundary of
      the Q-ball exits the horizon.
\item If we consider the scenario where the charges of the Q-balls
      evolve with time, the expectation value of the $\sigma$ in the
      core will also change with time. In this case we should
      examine when the transition from  
      $\sigma<\sigma_c$ to $\sigma>\sigma_c$ occurs.
      The properties of the Q-ball
      will be modified at $\sigma_c$. \footnote{Of course one might
      consider the scenario 
      where a huge Q-ball is produced at
      the earliest stage of the Universe and inflation starts abruptly. }
\item Naively, Q-balls might become a black hole before it induces
      inflation. This condition is rather trivial as we will show later.
\end{itemize}

Let us examine the above conditions in more detail.
First we assume $\eta\ll 1$, where both phases $\sigma<\sigma_c$ and
$\sigma>\sigma_c$ might appear.
When the charge $Q$ is small, the first term in (\ref{hybrid-potential})
dominates and  then (\ref{Q-ball-prop-gauge}) is reliable.\footnote{We are
not compelling the scenario where small Q-balls evolves with time.
Q-balls might be huge when it is produced before inflation.}
Since the vacuum energy $\rho$ inside Q-balls is as large as $M_0^4$, 
the Hubble constant when the boundary exits the horizon will be $H \simeq
\frac{M_0^2}{M_p}$.  
Then the condition  $r_Q > H^{-1}$ is represented by 
\begin{equation}
\label{gauge-inf}
Q^{1/4} \ge \frac{M_p}{M_0}.
\end{equation}
The above result can be applied as far as $\sigma <\sigma_c$.
However, the condition for $\sigma <\sigma_c$ is
\begin{eqnarray}
\label{gauge-equiv}
Q^{1/4} &<&  \left(\frac{M_0}{m_{3/2}}\right)\nonumber\\
&&=\left(\frac{M_p}{\Lambda_{SUSY}}\right)
\left(\frac{M_0}{\Lambda_{SUSY}}\right)\nonumber\\
&& = \frac{M_p}{M_0}\eta^2,
\end{eqnarray}
which is much smaller than (\ref{gauge-inf}).
Thus in this case, we must conclude that evolving Q-balls will alter
their properties from (\ref{Q-ball-prop-gauge}) to
(\ref{Q-ball-prop-gra}) before it induces inflation.
Of course the evolving Q-balls must not decay into black holes before it
induces inflation.
Denoting the Schwarzschild radius by $r_g$, one can easily find the condition
\begin{equation}
\label{gauge-BH}
Q^{1/4} \le \frac{M_p}{M_0} 
\end{equation}
which suggests that the critical charge for the black hole formation is
 much larger than the 
criteria (\ref{gauge-equiv}).\footnote{To understand (\ref{gauge-BH}),
we think it is helpful to consider a simplest toy model.
Imagine a spherical region with the radius $r_b$.
The vacuum energy is $\rho=\rho_b>0$ inside the ``ball'', while $\rho=0$
outside. 
Then the mass of the ``ball'' is given by the formula
$M_b=\frac{4\pi}{3}r_b^3 \rho_b$.
The Schwarzschild radius of the ``ball'' is $r_g^2 \simeq \rho_b/M_p^2$.
On the other hand, the Hubble parameter inside the ``ball'' when the
 boundary exits horizon is $H^2 \simeq \rho_b/M_p^2$.
Thus the naive black hole condition will give a rather trivial result.}

Now the above results are suggesting that we should consider the Q-balls of
$\sigma > \sigma_c$ in the case when $\eta \ll 1$.
When the Q-ball become large and the expectation value of the field
$\sigma$ inside the Q-ball becomes larger than 
$\sigma_c$, the above criteria are represented by the 
following conditions.
Here the vacuum energy density inside Q-balls is estimated by
$\rho \simeq m_{3/2}^2 \left[(|K|^{3/4})m_{3/2}Q^{1/2}\right]^2$.
The radius of the Q-ball will exit the horizon when $r_Q > H^{-1}$, which
happens when
\begin{eqnarray}
\label{gra-inf}
Q &>& |K_2|^{-1/2}\left(\frac{M_p}{m_{3/2}}\right)^2\nonumber\\
&&=|K_2|^{-1/2}\left(\frac{M_p}{\Lambda_{SUSY}}\right)^4\nonumber\\
&&=|K_2|^{-1/2}\left(\frac{M_p}{M_0}\right)^4\times \eta^4,
\end{eqnarray}
To verify our argument, we should examine if $\sigma >\sigma_c$ is
satisfied when inflation starts.
$\sigma >\sigma_c$ is satisfied if
\begin{eqnarray}
\label{gra-equiv}
Q &>&  |K_2|^{-3/2} 
\left(\frac{M_p}{m_{3/2}}\right)^2 \times \eta^4 \nonumber\\
&&=|K_2|^{-3/2} 
\left(\frac{M_p}{M_0}\right)^4 \times \eta^8.
\end{eqnarray}
Because we are considering the case $\eta \ll 1$, the condition
(\ref{gra-equiv}) is safely satisfied when inflation starts. 

Now we will examine the case $\eta \gg 1$.
As we are considering hybrid potential in this paper, $\eta \gg 1$
is not unnatural.
As we have discussed above, we should always use (\ref{Q-ball-prop-gauge}),
because in this case $\sigma_c$ is much larger than $M_p$.
The conditions for $r_Q > H^{-1}$ and $r_Q > r_g$ are the same as 
the result obtained above.
The properties of the Q-balls do not alter as the charge evolves, which 
is the only difference from the above result.
In this case, $\sigma\ll \sigma_c$ is always satisfied for any realistic
value of $Q$. 
Thus our conclusion for $\eta \gg 1$ is that the Q-ball inflation
will start 
when $Q \ge (M_p/M_0)^4$.

In the above discussions we have examined the conditions for inflation
to start within Q-balls.
Our second task is to examine the evolution of the field $\sigma$ during
the Q-ball inflation.
At the earliest stage of the Q-ball inflation, the field $\sigma$ is
trapped at the false vacuum because of the large $\omega$.
When inflation starts inside the Q-ball, the friction term dominates the
equation of motion.
Then $\dot{\sigma}$ decays as $\dot{\sigma}\sim e^{Ht}$.
Assuming that the change in $|\sigma|$ is much slower than that in
$\omega$, one can obtain 
\begin{equation}
\label{omega}
\omega\simeq \omega_0 e^{Ht},
\end{equation}
where $\omega_0$ denotes the initial value of $\omega$ when inflation
starts.
In this case, one may expect two kinds of inflation that may start
subsequently.
The first inflation occurs during the period of $\omega^2 > V''$ when the field
$\sigma$ is trapped at the false vacuum because of the large $\omega$.
The e-foldings of the first inflation is 
\begin{equation}
N_e \simeq \frac{1}{2}\log \left(\frac{w_0^2}{V''}\right),
\end{equation}
where $V''$ is the effective mass of the field $\sigma$ at the false vacuum.
Then after this period, the conventional slow-roll inflation will start
if some conditions are satisfied.
If $\eta >1$, the Hubble parameter might be smaller than the effective
mass, where the situation is the same as the conventional D-term
inflation. 
The e-foldings of the second inflation is determined by the charge of
the Q-ball, which determines the initial value of $\sigma$.
Even if the slow-roll inflation does not start, the expansion of the
local area induced by the small inflation will affect the later
cosmological structure formation. 
In any case, the cosmological observations of the present Universe might
put some bound on the Q-ball
inflation, which might put a bound on the phenomonological models.

Our last example is the Randall-Sundrum Type 2(RS-2) model\cite{RS2}.
In this case the Friedman equation receives an additional term that is
quadratic in the density.
The Hubble parameter is related to the energy density by
\begin{equation}
H^2 = \frac{8\pi}{3M_p^2}\rho \left(1+\frac{\rho}{2\lambda}\right),
\end{equation}
where $\lambda$ is the brane tension.
Denoting the ratio between $\rho$ and $\lambda$ by $\epsilon\equiv
\lambda/\rho$, the modified condition for $H^{-1}<r_Q$ is relaxed when
$\epsilon <1$.
Assuming that $\epsilon \ll 1$, one can obtain for $\sigma<\sigma_c$,
\begin{equation}
\label{RS-inf}
Q \ge \left(\frac{M_p}{M_0} \right)^4\times \epsilon^2.
\end{equation}
For $\sigma> \sigma_c$, it becomes
\begin{equation}
\label{RS-inf2}
Q >|K_2|^{-1/2}\left(\frac{M_p}{M_0}\right)^4 \times \eta ^4 \epsilon.
\end{equation}
From (\ref{RS-inf}) and (\ref{RS-inf2}), one can see that the condition
for Q-ball inflation is rather relaxed in the
Randall-Sundrum Type 2 scenario.

\section{Surface reheating and Q-ball inflation}
\hspace*{\parindent}
In this section we will examine whether the Q-balls produced just after
inflation leads to the surface reheating or to the Q-ball inflation.
The surface reheating in models of running mass inflation is already
examined in ref.\cite{surface-reheat}.
Here we mainly follow the setups of \cite{surface-reheat}, and discuss
if the Q-ball inflation takes place.
The potential which can lead to a Q-ball formation is 
\begin{equation}
\label{potential_running}
V(\sigma)=m^2 \left(1+ |K| \log \left[\frac{\sigma^2}{M_p^2}\right]\right)
\end{equation}
We will assume that the maximum charge of a Q-ball might be as large as the
maximum charge within the Hubble radius when Q-balls are produced.
Denoting the Hubble parameter and the expectation value of the field
$\sigma$ at the time of the Q-ball formation by $H_Q$ and $\sigma_Q$,
one can obtain the maximum charge $Q_{MAX}$,
\begin{equation}
\label{Q-Max}
Q_{MAX}\simeq \frac{\omega_Q \sigma_Q^2}{H_Q^3}.
\end{equation}
In generic situations, $\omega_Q$ is expected to be about the same
order as the Hubble parameter $H_Q$.
Finally, one can obtain the simple result,
\begin{equation}
Q_{MAX} \simeq \sqrt{K} \left(\frac{M_p}{m}\right)^2.
\end{equation}
which is nearly the same order as
(\ref{gra-inf}).\footnote{Note that $m$ in (\ref{Q-Max}) corresponds to 
$m_{3/2}$ in (\ref{gra-inf}).} 
However, it becomes much smaller than (\ref{gra-inf}) if $K \ll 1$.
Although it is quite difficult to calculate the exact values of the 
properties of the cosmological Q-balls, it seems fair to conclude from
the above 
rough estimations that the surface reheating will be reliable if $K$ is
much smaller than $O(1)$.\footnote{This result does not exclude 
the possibility that a few Q-balls might become abnormally large 
absorbing other Q-balls.}

\section{Conclusions and Discussions}
\hspace*{\parindent}
In this paper we have examined the idea that a modified version of
the conventional topological inflation might start within Q-balls.
An extension of the Q-balls is discussed by using an explicit hybrid
potential, which is useful for our discussion.
As the flat direction might appear in the hybrid potential 
of the brane distance, Q-balls might appear for the brane
rotation\cite{incidental_matsuda}.
We have obtained the criteria for the Q-ball inflation, which is
comparable to the phenomenological values of the conventional Q-ball
formation. 
Since the Q-ball formation is quite general in many models where flat
directions are contained in the low-energy effective Lagrangian,
we think the cosmological bound that will be obtained by considering
Q-ball inflation is quite important.

\section{Acknowledgment}
We wish to thank K.Shima for encouragement, and our colleagues in
Tokyo University for their kind hospitality.

\end{document}